# Charging and Growth of Fractal Dust Grains

Lorin S. Matthews and Truell W. Hyde, *Member, IEEE*

*Abstract*— The structure and evolution of aggregate grains formed within a plasma environment are dependent upon the charge acquired by the micron-sized dust grains during the coagulation process. The manner in which the charge is arranged on developing irregular structures can affect the fractal dimension of aggregates formed during collisions, which in turn influences the coagulation rate and size evolution of the dust within the plasma cloud. This paper presents preliminary models for the charge and size evolution of fractal aggregates immersed in a plasma environment calculated using a modification to the orbital-motion-limited (OML) theory.

Primary electron and ion currents incident on points on the aggregate surface are determined using a line-of-sight (LOS) approximation: only those electron or ion trajectories which are not blocked by another grain within the aggregate contribute to the charging current. Using a self-consistent iterative approach, the equilibrium charge and dipole moment are calculated for the dust aggregate. The charges are then used to develop a heuristic charging scheme which can be implemented in coagulation models. While most coagulation theories assume that it is difficult for like-charged grains to coagulate, the OML_LOS approximation indicates that the electric potentials of aggregate structures are often reduced enough to allow significant coagulation to occur.

## I. INTRODUCTION

AN understanding of the physics behind the coagulation of micron-sized dust grains is essential to research across fields as diverse as planetesimal formation and the plasma processing of silicon wafers for computer chips. The charge collected on such micron-sized dust grains when immersed in a plasma plays a crucial role in determining the overall structure and evolution of the aggregates resulting from coagulation of the grains. In particular, the manner in which the charge is arranged on the grain due to anisotropic charging currents, charge rearrangement on the surface, or induced charge-dipole interactions can be a critical factor in determining both the coagulation rate and the overall size evolution of the dust cloud [1].

While the charge on a spherical dust grain immersed in a plasma environment is fairly easy to determine under the proper set of assumptions [2, 3], calculating the charge on a non-spherical object is much more complex. Astronomical and experimental data have both shown that dust grains most likely begin the coagulation process through the formation of fluffy fractal-like aggregates with the charge on these aggregates necessarily distributed over their entire surface [4, 5]. This charge distribution is of importance when modeling the interaction of two charged aggregates: not only does the electrostatic interaction influence whether or not aggregates will collide and stick, the deflection of impinging grains' trajectories due to locally non-isotropic electric fields helps determine whether the resultant structure is compact (high fractal dimension) or more open (low fractal dimension). This result is of vital important to the overall coagulation process since fractal aggregates exhibit stronger gas-grain coupling and have greater collisional cross sections due to their open nature.

Although any increase in collisional cross-sectional area increases the overall coagulation rate, strong gas-coupling reduces it through suppression of the relative velocities between aggregates. Thus an accurate understanding of the physical geometry of the forming system, the physics behind the underlying charging process and the manner in which the growing aggregate interacts with the ambient plasma environment are all essential to properly modeling dust coagulation. Since both coagulation and charging are computationally intensive to model for fractal aggregates, it is desirable to develop a heuristic charging scheme from a detailed charging model, which can then be incorporated into a coagulation model.

This study presents such a model for the charging of fractal aggregates, formed from the coagulation of spherical monomers, using a modified orbital-motion-limited (OML) theory. The charge on the aggregates is approximated by a multipole expansion with the charging currents to points on the aggregate calculated using a line-of-sight (LOS) approximation to allow for the grain's morphology. The results obtained are compared to the charges calculated for growing fractal aggregates assuming charge conservation during collisions between monomers or aggregates [6].

## II. METHOD

The surface potential of single, small isolated grains is most often calculated using OML theory. This theory is based on the assumption that energy and momentum are conserved for impinging current species and that ions and electrons which have encountered potential barriers (such as those where their trajectories intersect another grain) have been removed from the background Maxwellian distribution [3, 7, 8]. As a result, this method of calculation becomes more difficult for non-spherical grains. In this work, a line-of-sight approximation is employed for aggregates built from spherical monomers in order to determine the locations on the surface of each monomer which are not blocked by other monomers within the aggregate.

### A. Modified Orbital Motion Limited Theory

The equilibrium charge on a grain can be determined once the sum of the currents to the grain are zero. The current density of species α (usually electrons or ions) to any point on a grain is given by

$$J = n_{\alpha\infty} q_\alpha \int f v \cos\theta\, d^3\vec{v} \qquad (1)$$

where $n_{\alpha\infty}$ is the number density outside the grain's potential well, $q_\alpha$ is the charge on the plasma species, $f$ is the distribution function, $\theta$ is the angle at which an orbit or path intersects the surface, and the integration is carried out over the velocity space $d^3\vec{v}$ of all orbits which intersect the surface for the first time. Assuming a Maxwellian distribution for the ambient plasma, the distribution function is given by

$$f = \left(\frac{m_\alpha}{2\pi k T_\alpha}\right)^{3/2} \exp\left(-\frac{m_\alpha}{2k T_\alpha} v^2 - \frac{q_\alpha \phi}{k T_\alpha}\right) \qquad (2)$$

where $m_\alpha$ and $T_\alpha$ are the mass and temperature of the plasma species, respectively, k is Boltzmann's constant, and $\phi$ is the grain potential. Making the substitution

$$d^3\vec{v} = v^2 d^2\Omega\, dv, \qquad (3)$$

where $d^2\Omega$ is the differential solid angle, allows the integration over the velocity to be separated from the integral over the angles. The limits of integration over the velocity run from $v_{min}$ to infinity, where $v_{min}$ is determined by requiring conservation of energy:

$$v_{min} = \begin{cases} 0 & q\phi \geq 0 \\ \sqrt{-\dfrac{2q\phi}{m}} & q\phi < 0 \end{cases} \qquad (4)$$

### B. Line of Sight Approximation

Given the above, calculating the charging on a fractal aggregate is reduced to determining the solid angle, $d^2\Omega$, for unobstructed orbits to each monomer within the aggregate. In this work, a line-of-sight approximation was used to determine which orbits would be removed from the integration over the distribution function [9]. It is assumed that electrons or ions move in a straight line from infinity and that the flux to any point on the surface of a monomer from a direction whose line of sight is blocked by a grain, including itself, is excluded from the integration in Eq. (1) (see Fig. 1). It is important to note that this approximation becomes less accurate as a grain becomes more highly charged since in this case ion orbits can have significant curvature in the vicinity of the charged grain. After calculating the LOS factor, $d^2\Omega$, separately for each constituent monomer, the integration in Eq. (1) is then carried out over the entire surface of the aggregate.

Assuming that the aggregate consists of a dielectric material such as water ice, a material commonly found in astrophysical environments, the impinging charged species will remain near the point of impact. This allows the monopole and dipole contributions to the potential to be calculated for individual monomers, with the sum of these determining the monopole and dipole potential of the entire aggregate. This new potential is then used to recalculate the charging currents to the grains and the process repeated until the grains attain an equilibrium potential.

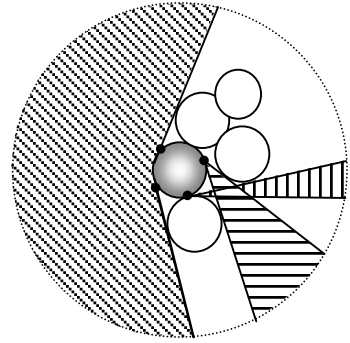

Fig. 1. The LOS factor $d^2\Omega$ is determined for each monomer by finding the unblocked trajectories of ions and electrons incident from infinity to given points on the monomer. The shaded regions on this 2D cross section indicate the contributions to the LOS factor from four points on a monomer's surface.

### C. Comparison of Charges on Aggregates

Charging calculations were carried out for aggregates formed from initial populations of grains with two different size distributions. The first consisted of grains ranging in size from 0.5-10 μm with an exponential size distribution, $n(a)da = a^\gamma da$, where $\gamma = -1.8$ and $n(a)\,da$ represents the number of particles with radii in the range [a, a+da]. The second population consisted of grains with an linear distribution in size and radii ranging from 1 μm to 6 μm.

The OML_LOS approximation was used to calculate the charge on fractal aggregates which were assembled starting from an initial seed monomer. The charge on the aggregate was calculated after the addition of each monomer or cluster of monomers. In a previous study, the initial monomers were charged to a potential of -1 V and the charge on the aggregate calculated assuming charge conservation during collision leading to a total charge $Q_o$ on the aggregate [6]. A dipole moment, $\mathbf{p}_o$, for the aggregate was then calculated assuming the charge was distributed over the entire aggregate structure, with a charge of

$$q_i = \frac{Q d_i}{\sum d_i} \qquad (5)$$

on the $i^{th}$ monomer, where $d_i$ is the distance of the $i^{th}$ monomer from the center of mass. Aggregates were grown in this manner to a maximum of one hundred monomers, where mutual electrostatic repulsion between particles made further coagulation growth unlikely.

In this study, a new equilibrium charge and dipole moment, $Q_{OML}$ and $\mathbf{p}_{OML}$, were determined for each aggregate using the OML_LOS approximation. The electron and ion temperatures used in the simulation were $T_e = T_i = 4637$ K, yielding a potential of -1 V on a spherical, isolated grain. The plasma number density was assumed to be $n_{\alpha\infty} = 5 \times 10^8$ m$^{-3}$, which is typical of astrophysical plasmas, although it is interesting to note that below a certain limit a change in density only affects the charging time and not the overall equilibrium potential. Charges were calculated for aggregates ranging in size from N = 2 to N ≈ 100, where N is the number of monomers in each aggregate. An exponential fit to the data allowed the OML charge and dipole moment to be calculated as a function of $Q_o$, $\mathbf{p}_o$, and N. This expression was then used as a heuristic charging algorithm to calculate the charge and dipole moment of aggregates as they were assembled from a single monomer as in [6]. The resulting charge and dipole moments for these aggregates, $Q_1$ and $\mathbf{p}_1$, were then compared to the values calculated employing OML_LOS.

III. RESULTS

The aggregate charge and dipole moment calculated using OML_LOS were found to be, in general, smaller than those calculated assuming charge conservation, a result which is to be expected for aggregates immersed in a plasma environment where grain charging is a continual process. Results comparing the charge and magnitude of the dipole moment are shown in Figs. 2 and 3 for the exponential size distribution; the results for the linear size distribution (Figs. 4 and 5) are similar.

As can be seen in Figs. 2 and 4, the charge ratio $Q_{OML}/Q_o$ is linearly decreasing on a log-log plot. A best-fit trend line gives the relation $Q_{OML} = 1.02$ N$^{-0.56}Q_o$ for the exponential size distribution and $Q_{OML} = 1.00$ N$^{-0.50}Q_o$ for the flat size distribution. Figures 3 and 5 show that the magnitude of the dipole moment is not strongly dependent on the number of monomers in the aggregate with $|\mathbf{p}_{OML}| = 0.079$ N$^{-0.07}$ $|\mathbf{p}_o|$ for the exponential size distribution and $|\mathbf{p}_{OML}| = 0.10$ N$^{-0.12}$ $|\mathbf{p}_o|$ for the flat distribution. There is large scatter in these correlations due to the wide variety of possible aggregate structures for any given number of monomers, since the structure plays a dominant role in the determination of the charge distribution and resulting dipole moment.

New aggregates were assembled following the method outlined in [6] using trend data from OML_LOS as a heuristic charging scheme to calculate the new charge and dipole moment, $Q_1$ and $\mathbf{p}_1$, on an aggregate after each collision.

The equilibrium charges and dipole moments of the new aggregate structures were also calculated directly employing LOS_OML theory. As seen in Figs. 2 -5, the ratios $Q_{OML}/Q_1$ and $|\mathbf{p}_{OML}|/|\mathbf{p}_1|$ are near unity across the range of aggregate sizes, indicating that the heuristic scheme for calculating grain charges is a valid approach. However, there is significant deviation for large N, with the heuristic scheme under-predicting the charge and dipole moment by a factor of two.

The fractal dimension of aggregate populations assembled employing charge-conservation and those using OML_LOS are compared for the exponential size distribution in Fig. 6. One striking difference is that the lower grain charges predicted by OML_LOS would allow for the formation of significantly larger aggregates, with lower fractal dimensions.

The fractal dimension for aggregates grown assuming charge conservation remains fairly constant as the aggregates increase in size, presumably because growth main occurs due to the addition of either monomers or very small clusters. Aggregates grown with the lower charges predicted by OML_LOS however, are able to not only grow much larger in size, but the fractal dimension is much lower due to growth as a results of cluster-cluster collisions.

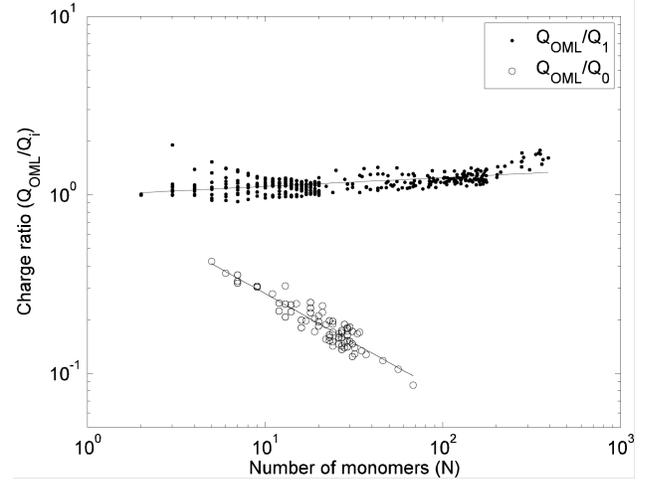

Fig. 2. Comparison of aggregate charge for the exponential size distribution. Open circles indicate the ratio of the charge calculated using OML_LOS to that calculated using charge conservation during collisions with the trend line showing the linear best fit. The points indicate the ratio of the charge calculated using OML_LOS to that calculated using the heuristic charging scheme during aggregate growth. The linear trend line for this dataset is close to unity, with most of the deviation at the endpoints.

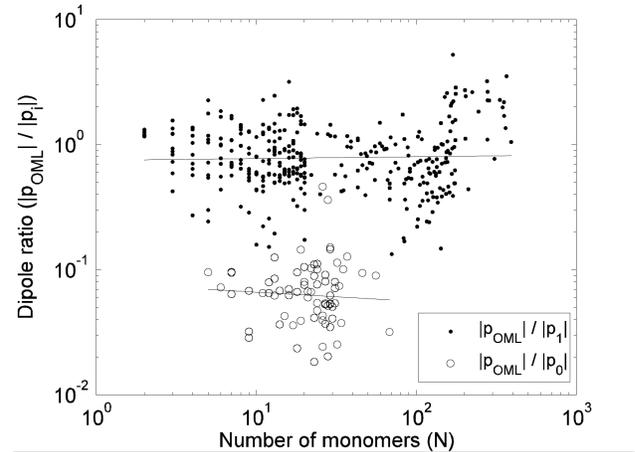

Fig. 3. Comparison of aggregate dipole moment for the exponential size distribution. Open circles indicate the ratio of the dipole moment calculated using OML_LOS to that calculated using charge conservation during collisions with the trend line showing the linear best fit. Although there is scatter in the data, on average the dipole moment calculated from OML_LOS is about 8% of that calculated using charge conservation. The points indicate the ratio of the dipole moment calculated using OML_LOS to that calculated using the trend line from the previous data set as a heuristic charging scheme. The linear trend line for this dataset shows an average near unity with large scatter.

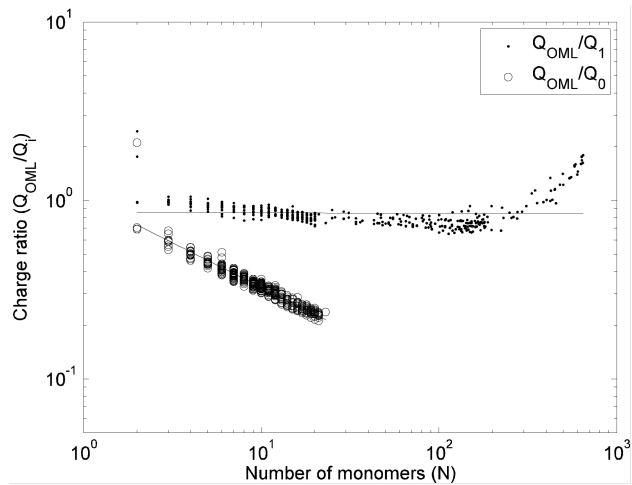

Fig. 4. Comparison of aggregate charge for the flat size distribution. Open circles indicate the ratio of the charge calculated using OML_LOS to that calculated using charge conservation during collisions with the trend line showing the linear best fit. The points indicate the ratio of the charge calculated using OML_LOS to that calculated using the heuristic charging scheme during aggregate growth. The linear trend line for this dataset is close to unity, with most of the deviation at the endpoints.

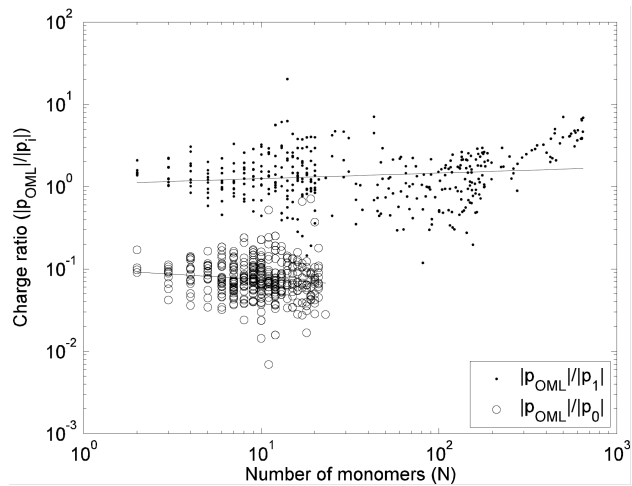

Fig. 5. Comparison of aggregate dipole moment for the flat size distribution. Open circles indicate the ratio of the dipole moment calculated using OML_LOS to that calculated using charge conservation during collisions with the trend line showing the linear best fit. Although there is scatter in the data, on average the dipole moment calculated from OML_LOS is about 8% of that calculated using charge conservation. The points indicate the ratio of the dipole moment calculated using OML_LOS to that calculated using the trend line from the previous data set as a heuristic charging scheme. The linear trend line for this dataset shows an average near unity with large scatter.

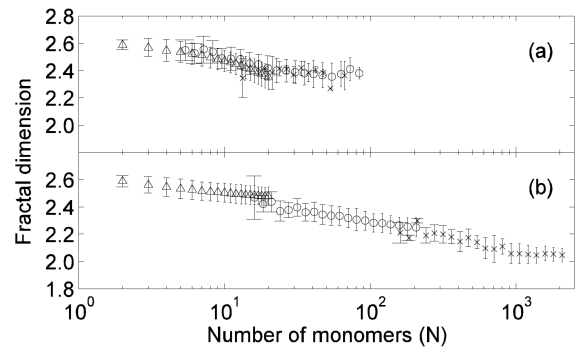

Fig. 6. Comparison of fractal aggregates built from an initial exponential size distribution using (a) charge conservation and (b) charges calculated from OML_LOS. The fractals were assembled in three generations (designated by triangles, circles and crosses respectively) by the addition of either monomers (first generation only) or aggregates from the previous generation(s). The primary difference between the two models is that OML_LOS predicts a smaller aggregate charge than charge conservation, allowing the formation of larger aggregates with a lower fractal dimension.

## IV. Conclusion and Discussion

A model for approximating the charge and dipole moment of an aggregate structure immersed in a plasma environment has been presented. The model calculates currents to each point on the aggregate surface using a LOS approximation to the OML theory: only those electron or ion trajectories which are not blocked by another grain within the aggregate contribute to the charging current. Preliminary data suggest that the overall charge on the aggregate structure can be well-approximated by a heuristic approach, with the charge being a function of the number of monomers in the aggregate, N, and the sum of the original charges on the constituent monomers, $Q_o$. The dipole moment of the aggregate structure is more difficult to predict as the geometry of the aggregate structure can vary greatly for a given N. The above suggests that a heuristic charging scheme is desirable for implementation of rapid charge calculations in an N-body coagulation model.

In general, the charge and dipole moments predicted by OML_LOS theory are smaller than those calculated assuming simple conservation of charge. This is to be expected in a plasma environment where the charging currents to the grains are continuous and an important factor in dust coagulation, as the reduced charges on the aggregate structure are less likely to inhibit coagulation.

Experimental data shows that aggregates grown by cluster-cluster aggregation (CCA) are generally non-spherical and have characteristic axial ratios ≈ 2.0 [10]. These aggregates tend to align during sedimentation along the drift axis, an effect that is observed in astrophysical environments by the subsequent polarization of transmitted light. In protoplanetary disks, this effect is greatest for large aggregates (N > 1000) where Brownian motion is no longer large enough to randomize grain orientation. This is very interesting since the reduced charge on the aggregates predicted by OML_LOS in this study may well be necessary for such CCA to occur and the subsequent charge arrangement and dipole interaction between the grains and their environment could play an important role in the overall alignment and structure formation of precursors to planetesimals.

Further work is needed to address the validity of the approximations made in the LOS_OML model. One primary concern is the assumption that ion trajectories will not deviate from a straight-line approach to an aggregate. In reality, ion trajectories can deviate substantially from a straight-line path, resulting in the ions impacting at locations normally hidden by a line-of-sight approximation. This is of greater importance when trying to determine the overall charge structure on a fractal aggregate and the resultant dipole moment.

It should also be noted that the electron and ion distributions in this study were assumed to be Maxwellian; space plasmas are more likely to have plasma distributions which are Lorentzian [11]. Under this type of distribution, grains can charge to a higher potential which may well play an important role in the coagulation of charged dust within protoplanetary disks leading to a change in planetary formation rates. Finally, the dust in a protoplanetary disk is immersed in a radiative as well as a plasma environment. Charging currents due to secondary electron emission could lead to positive grain charging, with different aggregates having potentials of opposite sign, further increasing coagulation rates. Another possibility is for aggregates consisting of dielectric materials to develop regions of both positive and negative charge, which could play an important role in the morphology of new aggregates built through collisions.

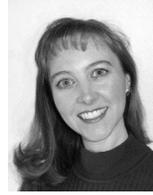

**Lorin S. Matthews** was born in Paris, TX in 1972. She received the B.S. and the Ph.D. degrees in physics from Baylor University in Waco, TX, in 1994 and 1998, respectively.

She is currently an Assistant Professor in the Physics Department at Baylor University. Previously, she worked at Raytheon Aircraft Integration Systems where she was the Lead Vibroacoustics Engineer on NASA's SOFIA (Stratospheric Observatory for Infrared Astronomy) project.

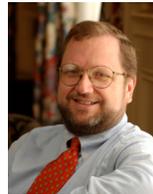

**Truell W. Hyde** was born in Lubbock, Texas in 1956. He received the B.S. in physics and mathematics from Southern Nazarene University in l978 and the Ph.D. in theoretical physics from Baylor University in 1988.

He is currently at Baylor University where he is the Director of the Center for Astrophysics, Space Physics & Engineering Research (CASPER), a Professor of physics and the Vice Provost for Research for the University. His research interests include space physics, shock physics and waves and nonlinear phenomena in complex (dusty) plasmas.